\documentclass[aps,twocolumn,showpacs,superscriptaddress,amsmath,amssymb,floatfix]{revtex4-1}

\usepackage[utf8]{inputenc}
\usepackage{siunitx}
\usepackage{graphicx}
\usepackage{color}
\usepackage[english]{babel}
\usepackage{subscript}
\usepackage{xspace}
\usepackage{hyperref}
	\definecolor{LinkColor}{rgb}{0.45,0,0}
	\definecolor{UrlColor}{rgb}{0,0,0.45}
	\definecolor{CiteColor}{rgb}{0,0.45,0}
\hypersetup{%
	linkcolor=LinkColor,%
	filecolor=LinkColor,%
	menucolor=LinkColor,%
	citecolor=CiteColor,%
	urlcolor=UrlColor,%
	colorlinks=true%
}

\begin{document}

\title{Enhanced ferrimagnetism in auxetic NiFe\textsubscript{2}O\textsubscript{4}\\in the crossover to the ultrathin film limit}

	\author{Michael Hoppe}
	\affiliation{Peter Gr\"unberg Institut (PGI-6) and
	JARA J\"ulich-Aachen Research Alliance, Forschungszentrum J\"ulich GmbH, 52425 J\"ulich, Germany}

	\author{Sven Döring}
	\affiliation{Peter Gr\"unberg Institut (PGI-6) and
JARA J\"ulich-Aachen Research Alliance, Forschungszentrum J\"ulich GmbH, 52425 J\"ulich, Germany}

	\author{Mihaela Gorgoi}
	\affiliation{Helmholtz-Zentrum Berlin für Materialien und Energie GmbH, 12489 Berlin, Germany}

	\author{Stefan Cramm}
	\affiliation{Peter Gr\"unberg Institut (PGI-6) and
JARA J\"ulich-Aachen Research Alliance, Forschungszentrum J\"ulich GmbH, 52425 J\"ulich, Germany}

	\author{Martina M\"uller}
	\email{mart.mueller@fz-juelich.de}
	\affiliation{Peter Gr\"unberg Institut (PGI-6) and
JARA J\"ulich-Aachen Research Alliance, Forschungszentrum J\"ulich GmbH, 52425 J\"ulich, Germany}
	\affiliation{Fakult\"at f\"ur Physik, Universit\"at Duisburg-Essen, 47048 Duisburg, Germany}

	\date{\today{}}


%%%%%%%%%%%%%%%%%%%%%%%%%%%%%%%
%	Abstract                  %
%%%%%%%%%%%%%%%%%%%%%%%%%%%%%%%
\begin{abstract}
	We investigate the sensitive interplay between magnetic, electronic and structural properties in the ferrimagnetic oxide NiFe\textsubscript{2}O\textsubscript{4}. Emphasis is placed on the impact of reduced dimensionality in the crossover from bulk-like to ultrathin films. We observed an enhanced saturation magnetization $M_S$ for ultrathin NiFe\textsubscript{2}O\textsubscript{4} films on Nb-SrTiO\textsubscript{3} (001) substrates that co-occurs with a reduced out-of-plane lattice constant under compressive in-plane epitaxial strain. We found a bulk-like cationic coordination of the inverse spinel lattice independent of the NiFe\textsubscript{2}O\textsubscript{4} film thickness -- thus ruling out a cationic inversion that nominally could account for an enhanced $M_S$. Our study instead uncovers a reduction of the unit cell volume, i.e. an auxetic behavior in ultrathin NiFe\textsubscript{2}O\textsubscript{4} films, which may result in an enhanced magnetic exchange caused by an increased interatomic electronic localization.
\end{abstract}

\maketitle

%%%%%%%%%%%%%%%%%%%%%%%%%%%%%%%%%%%%%%%%%%%
%	INTRODUCTION                          %
%%%%%%%%%%%%%%%%%%%%%%%%%%%%%%%%%%%%%%%%%%%
\section{Introduction}
\label{sec:introduction}

	The competition of charge, spin and orbital degrees of freedom in complex oxides leads to intriguing physical phenomena, including ferromagnetism, ferroelectri\-city or multiferroi\-city \cite{hwang_emergent_2012}.  Fertilized by the continuously advancing art of oxide growth, the controlled synthesis of high-quality oxide heterostructures now approaches a monolayer-precision \cite{seshadri_advances_2012}. Designing electronic properties in ultrathin oxide films and interfaces thereby opens up routes to explore novel nanoelectronic functionalities for applications.

	In the context of spin-based electronics, oxides featuring both magnetic and insulating properties reveal a highly effective spin filter effect, where spin-polarized electron currents are generated by a spin-dependent tunnelling process. Up to \SI{100}{\percent} spin filtering has been demonstrated in magnetic oxides with low Curie temperature $T_c$, such as the binary rare earth compounds EuO or EuS \cite{miao_magnetoresistance_2009,muller_observation_2011}, and hence their integration as model spin injection/detection contacts to silicon was explored recently \cite{caspers_chemical_2011,caspers_heteroepitaxy_2013}. Implementing the spin filter functionality of magnetic insulators in all-oxide heterostructures can extend the scope of applications further towards a multifunctional oxide-based spin electronics.

	It is in this pursuit that ferrite materials are envisioned as high-$T_c$ spin filters with the ultimate goal to realize efficient spin filtering for application at room temperature. For example, NiFe\textsubscript{2}O\textsubscript{4} shows ferrimagnetic ordering up to $T_C = \SI{865}{\kelvin}$ \cite{v._a._m._brabers__1995} and grows epitaxially on Nb-doped SrTiO\textsubscript{3} (001) perovskite electrodes \cite{ma_robust_2010,hoppe_wide-range_2014}. Its inverse spinel lattice of the type Fe\textsuperscript{3+}[Ni\textsuperscript{2+}Fe\textsuperscript{3+}]O\textsubscript{4} however, exhibits a high structural complexity:  Ni\textsuperscript{2+}-cations are situated on octahedrally ($O_h$) coordinated lattice sites, while Fe\textsuperscript{3+}-cations are equally distributed across both tetrahedral ($T_d$) and $O_h$ sites (Fig. 1). The electronic and magnetic properties of spinel ferrites thus sensitively depend on the details of the interatomic coordinations. In particular, magnetic ordering is dominated by superexchange interactions between $T_d$ and $O_h$-coordinated cations on two antiferromagnetically coupled sublattices.

	\begin{figure}
	\includegraphics{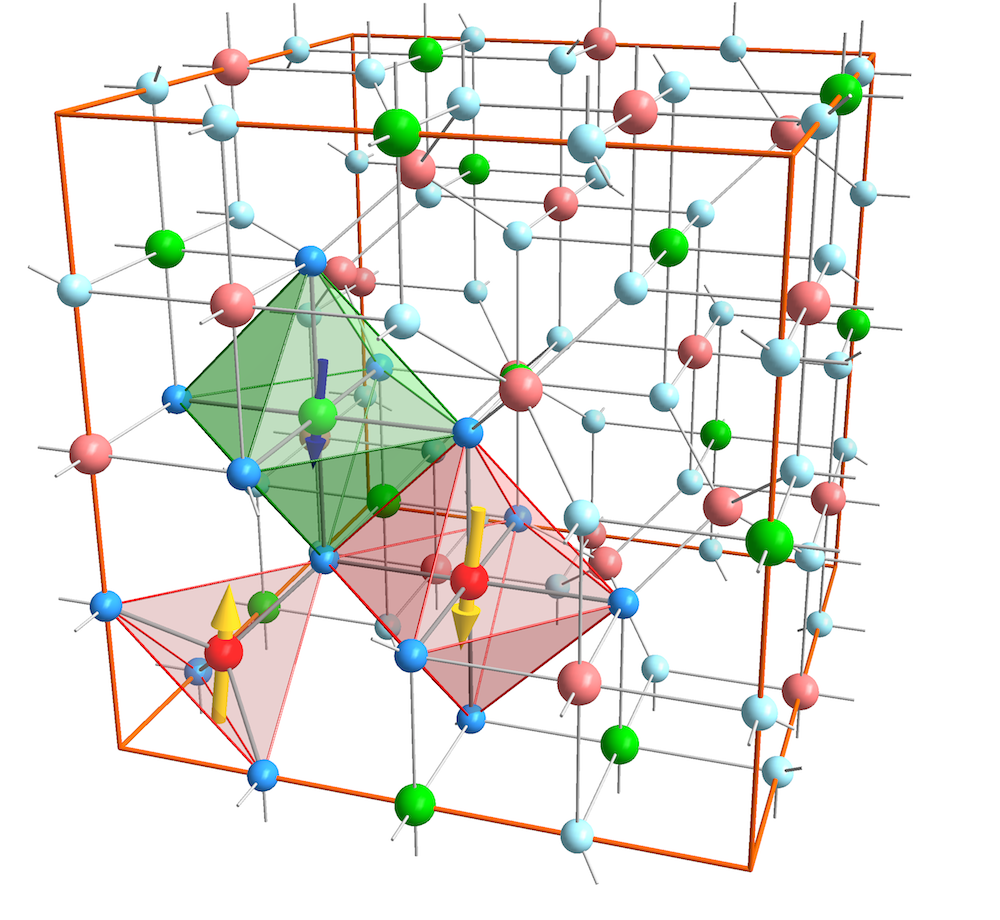}
	\vspace*{-6mm}
	\caption{Schematic representation of the inverse spinel lattice of NiFe\textsubscript{2}O\textsubscript{4}: Fe\textsuperscript{3+}-cations (red) are distributed equally across tetra- ($T_d$) and octahedral ($O_h$) lattice sites, while  Ni\textsuperscript{2+}-cations (green) occupy $O_h$ sites. An antiferromagnetic coupling between the $T_d$ and $O_h$ sites compensates the magnetic moments of the Fe\textsuperscript{3+}-cations, why only the Ni\textsuperscript{2+}-cations account for the net macroscopic magnetization of \mbox{2\,$\mu_B$/f.u..}}
	\label{fig:inv_spinel}
	\end{figure}

	A structural inversion from the inverse to the normal spinel lattice consequently alters the cationic coordination, as quantified by the inversion parameter $\lambda$. Hereby, $\lambda$ is the fraction of A\textsuperscript{2+}-cations occupying $O_h$ sites, with $\lambda = 0$ denoting a normal (Ni\textsuperscript{2+}[Fe\textsuperscript{3+}Fe\textsuperscript{3+}]O\textsubscript{4}) and $\lambda = 1$ an inverse (Fe\textsuperscript{3+}[Ni\textsuperscript{2+}Fe\textsuperscript{3+}]O\textsubscript{4}) spinel lattice. In previous studies, an unexpected magnetic behaviour, i.e. an enhanced saturation magnetization was reported for NiFe\textsubscript{2}O\textsubscript{4} films in the ultrathin film limit \cite{luders_enhanced_2005,venzke_epitaxial_1996}. The origin of this phenomenon was explained by a cationic inversion from an inverse to a partly normal spinel lattice ($0< \lambda < 1$), since this structural redistribution of Fe cations nominally accounts for an increased magnetic moment. Theoretical considerations based on density functional theory calculations find a partial cationic inversion energetically favourable for NiFe\textsubscript{2}O\textsubscript{4} films under tensile, but not under compressive strain \cite{fritsch_effect_2011}, as is the case for NiFe\textsubscript{2}O\textsubscript{4} grown on Nb-SrTiO\textsubscript{3} (001). The origin of the altered magnetic exchange interaction in ultrathin NiFe\textsubscript{2}O\textsubscript{4} films thus still remains an open question.

	In this work, we explore the details of the electronic and magnetic properties of single-crystalline NiFe\textsubscript{2}O\textsubscript{4} films in the crossover from bulk-like to the ultrathin film limit. The goal of our studies is to uncover modifications of the structural, electronic and magnetic properties with regard to the reduced film dimensionality.  We performed a complementing spectroscopic analysis employing the bulk- and surface sensitive photon spectroscopy techniques HAXPES, XANES and XMCD, respectively, which allow for a precise quantification of the element-specific cationic valencies and spatial coordinations. From our throughout analysis, we can conclude on the absence of a cationic inversion for all NiFe\textsubscript{2}O\textsubscript{4} film thicknesses. Instead, we propose an auxetic behaviour of ultrathin NiFe\textsubscript{2}O\textsubscript{4} films as the mechanism that may lead to an enhanced interatomic electronic localization.

%%%%%%%%%%%%%%%%%%%%%%%%%%%%%%%%%%%%%%%%%%%
%	EXPERIMENTAL                          %
%%%%%%%%%%%%%%%%%%%%%%%%%%%%%%%%%%%%%%%%%%%
\section{Experimental details}
\label{sec:experimental}

	A series of NiFe\textsubscript{2}O\textsubscript{4} thin films with varying thicknesses between 2 and \SI{20}{\nano\metre} has been deposited from stoichiometric targets on conductive \SI{0.1}{\percent} Nb-doped \mbox{SrTiO\textsubscript{3}(001)} substrates by utilization of the pulsed laser deposition technique. The substrates were previously etched in buffered hydrofluoric acid to provide a TiO\textsubscript{2}-terminated terrace surface \cite{koster_quasi-ideal_1998}. During growth, the laser fluence was set to \SI{1.5}{\joule\per\centi\metre^2} and a repetition rate of \SI{2}{\hertz}. The oxygen pressure was kept at \SI{0.04}{\milli\bar} and the substrate was heated to $T_S = \SI{635}{\celsius}$. After deposition, the samples were post-annealed at $T_S$ for \SI{90}{\minute} in vacuum. 

	The thickness of the grown films was determined by X-ray reflectivity measurements (XRR), while the structural characterization was accomplished by X-ray diffraction experiments (XRD). Both XRR and XRD experiments were performed on a Philips XPert MRD using Cu-$K\alpha$-radiation. Bulk magnetic properties of the samples were investigated on a Quantum Design MPMS SQUID. Hysteresis loops were recorded at $T = \SI{5}{\kelvin}$ with a magnetic field up to \SI{3.6}{\tesla}, which was applied parallel to the in-plane [100]-axis of the films.

	Hard X-ray photoelectron spectroscopy (HAXPES) was conducted on the HIKE endstation of the KMC-1 beamline at the BESSY-II electron storage ring (HZB Berlin) \cite{gorgoi_high_2009}. In contrast to soft X-ray photoelectron spectroscopy, HAXPES experiments use high energy X-rays with photon energies $E_P$ ranging from 2 to \SI{15}{\kilo\electronvolt}. Thus, the kinetic energy as well as the inelastic mean free path of the emitted photoelectrons are strongly enhanced, which gives HAXPES an information depth (ID) of several tens of nanometers, allowing one to probe the chemical properties of a multi-layered film structure with true bulk sensitivity \cite{drube_photoelectron_2005}. All spectra shown in this work were taken at $E_P = \SI{4}{\kilo\electronvolt}$ and at room temperature. The X-ray beam was aligned at \SI{3}{\degree} grazing incidence and the photoelectron detector normal to the sample surface. By defining ID(95) as the probing depth from which \SI{95}{\percent} of the photoelectrons originate, the given experimental parameters result in an ID(95) $\approx\SI{17}{\nano\metre}$. 

	X-ray absorption near edge structure (XANES) experiments have been performed at the HIKE endstation as well. To record the XANES spectra, the X-ray absorption of the samples was determined by the emitted fluorescence, which was detected by an energy dispersive detector. In contrast, the absorption of the NiFe\textsubscript{2}O\textsubscript{4} bulk target material was determined in total electron yield mode (TEY). All signals were normalized to the incident X-ray flux, monitored by a ionization chamber in front of the sample. The angle between the incident X-ray beam and the sample was optimized for every spectrum, in order to maximize the fluorescence signal without saturating the detector.

	For all photon energies used during the HAXPES and XANES experiments, photoemission spectra of the Au 4f core level from an Au reference sample attached to the manipulator have been recorded. The energy position of the Au 4f lines was compared to standard values and all measured data corrected accordingly.

	X-ray magnetic circular dichroism (XMCD) data was determined by X-ray absorption spectroscopy (XAS) experiments performed at the UE56-1 SGM beamline at BESSY-II. The sample surfaces were aligned in \SI{20}{\degree} grazing incidence. A magnetic field of 300 mT was applied parallel to the surface and in the plane spanned by the incident beam with the surface normal axis. The absorption signal was taken in TEY mode and was normalized to the incident X-ray flux. The XMCD asymmetry spectra were determined from the difference of two spectra collected by changing either the magnetic field to the opposite direction, or by two spectra recorded by changing the polarization from left to right-handed. In total, at least four absorption spectra were taken for every sample, each with a different combination of polarization and magnetization direction.

	For XMCD data analysis, we have calculated model XMCD spectra using the program CTM4XAS 5.5 \cite{stavitski_ctm4xas_2010}, which is based on crystal field multiplet calculations including charge transfer effects \cite{thole_spin-mixed_1988,ogasawara_theory_1991,ogasawara_praseodymium_1991}. Using the parameters from Ref.  \cite{pattrick_cation_2002}, the interatomic screening is taken into account by reducing the Slater integrals  F(dd), F(pd), and G(pd) with scaling factors F(dd) = 0.7, F(pd) = 0.8 and G(pd) = 0.8. For octahedral (tetrahedral) symmetry, the crystal field was set to 10 Dq = \SI{1.2}{\electronvolt} (\SI{-0.6}{\electronvolt}) and the exchange field was set to M = \SI{10}{\milli\electronvolt} (\SI{-10}{\milli\electronvolt}). The resulting spectra were broadened by a Lorentzian width with a half-width of 0.3 eV (0.5 eV) for the \textit{L}\textsubscript{3} (\textit{L}\textsubscript{2}) edge to respect the core-hole lifetime broadening, and by a Gaussian width of 0.2 eV to account for instrumental broadening.

%%%%%%%%%%%%%%%%%%%%%%%%%%%%%%%%%%%%%%%%%%%
%	RESULTS
%%%%%%%%%%%%%%%%%%%%%%%%%%%%%%%%%%%%%%%%%%%
\section{Results}
\label{sec:crystalline}

	\subsection{Structural and magnetic characterization}
	%%%%%%%%%%%%%%%%%%%%%%%%%%%%%%%%%%%%%%%%%%%
	%%%%%%%%%%%%%%%%%%%%%%%%%%%%%%%%%%%%%%%%%%%

	\begin{figure}
	\includegraphics[width=\columnwidth]{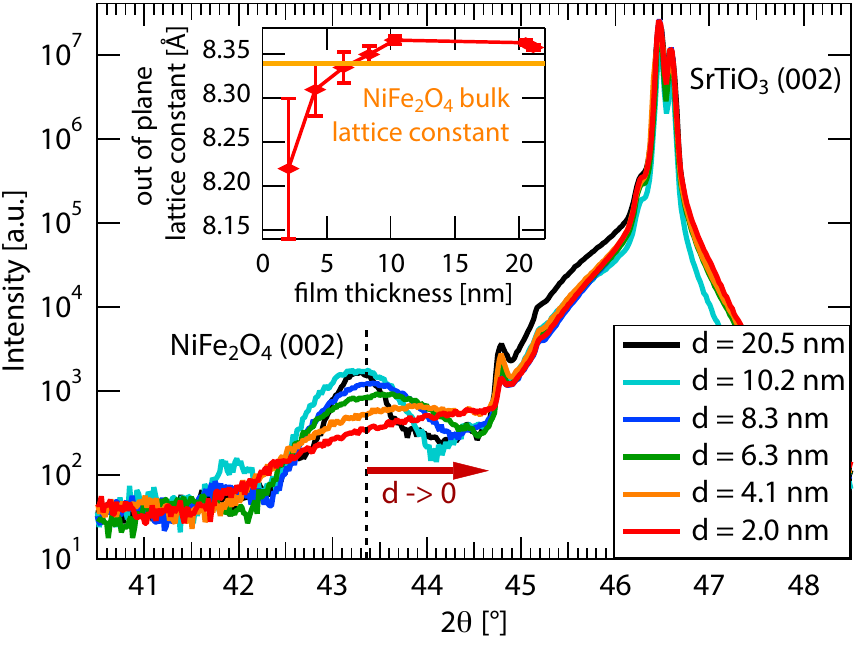}
	\vspace*{-6mm}
	\caption{$\theta$-$2\theta$-scans of the NiFe\textsubscript{2}O\textsubscript{4} (004) reflection for varying film thickness. The out-of-plane lattice constant $a_{\mathrm{oop}}$ decreases below the bulk value for ultrathin films (see inset).}
	\label{fig:xrd_thickness}
	\end{figure}

	First, the thickness-dependent structural properties of ultrathin NiFe\textsubscript{2}O\textsubscript{4} films on Nb-SrTiO\textsubscript{3} (001) were investigated by X-ray diffraction experiments. $\theta$-$2\theta$-scans ranging from $2\theta = \SI{20}{\degree}$ to $\SI{100}{\degree}$ with a scattering vector parallel to the surface normal were used to analyze the crystal structure of the films and to confirm their epitaxial growth. All scans show the expected reflections of a (001)-oriented SrTiO\textsubscript{3} crystal, as well as two additional reflections at $2\theta \approx \SI{43}{\degree}$ and $2\theta \approx \SI{95}{\degree}$, which can be attributed to the NiFe\textsubscript{2}O\textsubscript{4} (004) and (008) reflections. Since no other reflections are observed, we conclude that the NiFe\textsubscript{2}O\textsubscript{4} films grow textured along the (001) direction without any parasitic phases. $\Phi$-scans around the SrTiO\textsubscript{3} (202) and NiFe\textsubscript{2}O\textsubscript{4} (404) peaks both show a four-fold symmetry and provide evidence that the films grow cube-on-cube on the SrTiO\textsubscript{3} substrate, despite the induced biaxial compressive strain of \SI{6.4}{\percent}. In Figure \ref{fig:xrd_thickness}, the details of the $\theta$-$2\theta$-scans around the NiFe\textsubscript{2}O\textsubscript{4} (004) reflection are shown, which reveal that for decreasing film thickness the center of the NiFe\textsubscript{2}O\textsubscript{4} (004)-peak shifts towards larger angles, implying a decreasing out-of-plane lattice constant $a_{\mathrm{oop}}$. The broadening of the peaks for thinner films is due to the smaller amount of material that contributes to coherent diffraction. For film thicknesses above \SI{6}{\nano\metre}, $a_{\mathrm{oop}}$ is slightly larger than the bulk value (a\textsubscript{bulk} = \SI{8.339}{\angstrom}). In combination with the compressive in-plane stress induced by the substrate, this finding reveals the tendency of the material to preserve its bulk unit cell volume. On the other hand, for lower thicknesses $a_{\mathrm{oop}}$ decreases, as compiled in the inset of Fig. \ref{fig:xrd_thickness}. This refers to a reduction of the unit cell volume for ultrathin films in comparison to the bulk value, a result that also has been reported for CoFe\textsubscript{2}O\textsubscript{4} films on SrTiO\textsubscript{3} \cite{foerster_poisson_2012}. In contrast to CoFe\textsubscript{2}O\textsubscript{4}, however, $a_{\mathrm{oop}}$ of NiFe\textsubscript{2}O\textsubscript{4} even drops below its bulk value for ultrathin films, which implies that NiFe\textsubscript{2}O\textsubscript{4} shows an auxetic behaviour, i.e. a negative Poisson ratio $\nu$ in the crossover to the monolayer regime.

	\begin{figure}
	\includegraphics{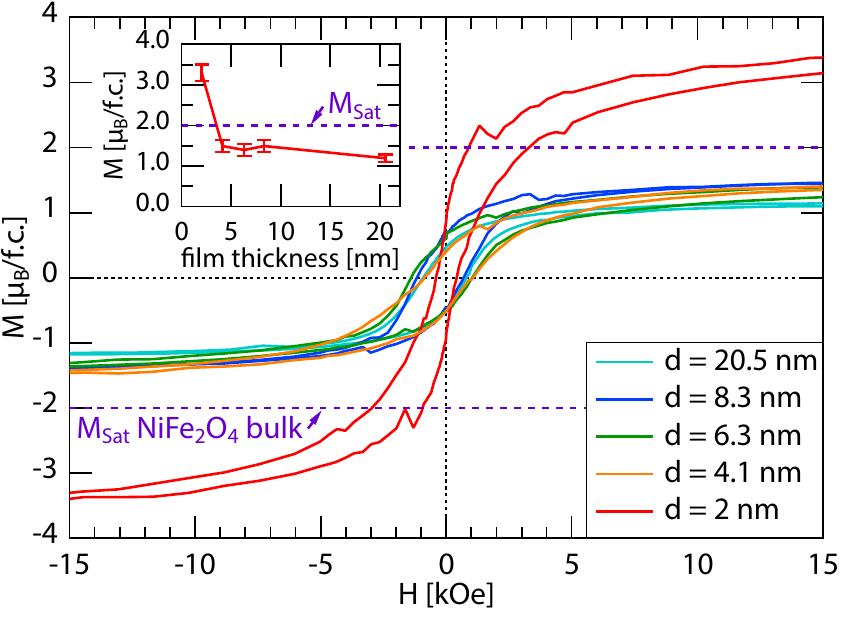}
	\vspace*{-6mm}
	\caption{In-plane $M$-$H$ hysteresis loops of NiFe\textsubscript{2}O\textsubscript{4} on SrTiO\textsubscript{3} (001) recorded at $T = \SI{5}{\kelvin}$. The inset shows the saturation magnetization $M_S$ as a function of NiFe\textsubscript{2}O\textsubscript{4} film thickness.}
	\label{fig:squid}
	\end{figure}

	Next, the NiFe\textsubscript{2}O\textsubscript{4} films were investigated with regard to their magnetic properties. Hereby, special attention is payed to changes dependent on their film thickness. Hysteresis loops of all samples were recorded at $T = \SI{5}{\kelvin}$, which are dominated by the diamagnetic contribution of the SrTiO\textsubscript{3} substrate. To extract the magnetic response of the NiFe\textsubscript{2}O\textsubscript{4} films, a subtraction of the diamagnetic background is required. Therefore, linear slopes have been fitted to the high-field tails of the raw signal and subtracted afterwards. 

	In Figure \ref{fig:squid}, hysteresis loops after background correction are depicted, which confirm ferromagnetic behaviour for all NiFe\textsubscript{2}O\textsubscript{4} film thicknesses. The coercive fields are approximately constant for thicknesses above $d = \SI{6}{\nano\metre}$, but dramatically decrease for $d = \SI{2}{\nano\metre}$. For CoFe\textsubscript{2}O\textsubscript{4} on MgO, this effect has been interpreted as a result of the reduction of the magnetic anisotropy for thin films  \cite{moyer_enhanced_2012}. NiFe\textsubscript{2}O\textsubscript{4} films with thicknesses above \SI{6}{\nano\metre} show a saturation magnetization of $M_S \approx 1.3 - 1.5\,\mu_B$/f.u., which is lower than the bulk value of $2\,\mu_B$/f.u. \cite{v._a._m._brabers__1995}. These deviations are supposed to be related to structural dislocations, which form due to the strain incorporated by the substrate, and to the formation of anti-phase boundaries during growth. The latter occur due to island forming at different positions on the substrate, which are shifted by half of a unit cell to each other and thus loose periodicity upon merging \cite{margulies_origin_1997}. This model is supported by the high external magnetic fields required to drive the films into saturation, which is even at $15\,$kOe not completely accomplished. More striking, when the film thickness scales below \SI{6}{\nano\metre}, we find the saturation magnetization enhancing up to 3 $\mu_B$/f.u. - thus significantly exceeding the bulk value. This result is in agreement with previous studies on NiFe\textsubscript{2}O\textsubscript{4}/SrTiO\textsubscript{3} \cite{luders_enhanced_2005} and CoFe\textsubscript{2}O\textsubscript{4}/SrTiO\textsubscript{3} \cite{rigato_magnetization_2009}. So far, this phenomenon was explained in terms of a cationic inversion, where the inverse spinel structure of the bulk state partly changes to a normal spinel structure in the crossover to the ultrathin film limit. An experimental proof for this model is however still lacking. 

	Moreover, for the thinner films, the contributions from contaminations to the total signal increase. Foerster \textit{et al.} \cite{foerster_distinct_2011} discussed the influence of the substrate, for which in the case of NiFe\textsubscript{2}O\textsubscript{4} films on MgAl\textsubscript{2}O\textsubscript{4}, the observed increased magnetization can be explained by a paramagnetic contribution from the substrate, which even disappears, if the magnetic response of the substrate is subtracted properly. Yet this cannot explain the findings for NiFe\textsubscript{2}O\textsubscript{4} on SrTiO\textsubscript{3}, since SrTiO\textsubscript{3} shows a purely diamagnetic response and thus validates the applied background subtraction. 

	In order to evidence the existence or absence of a cationic inversion, we investigate the chemical properties and cationic distribution of NiFe\textsubscript{2}O\textsubscript{4} as a function of the film thickness in more detail.

	\subsection{HAXPES}
	%%%%%%%%%%%%%%%%%%%%%%%%%%%%%%%%%%%%%%%%%%%
	%%%%%%%%%%%%%%%%%%%%%%%%%%%%%%%%%%%%%%%%%%%

	\begin{figure}
	\includegraphics{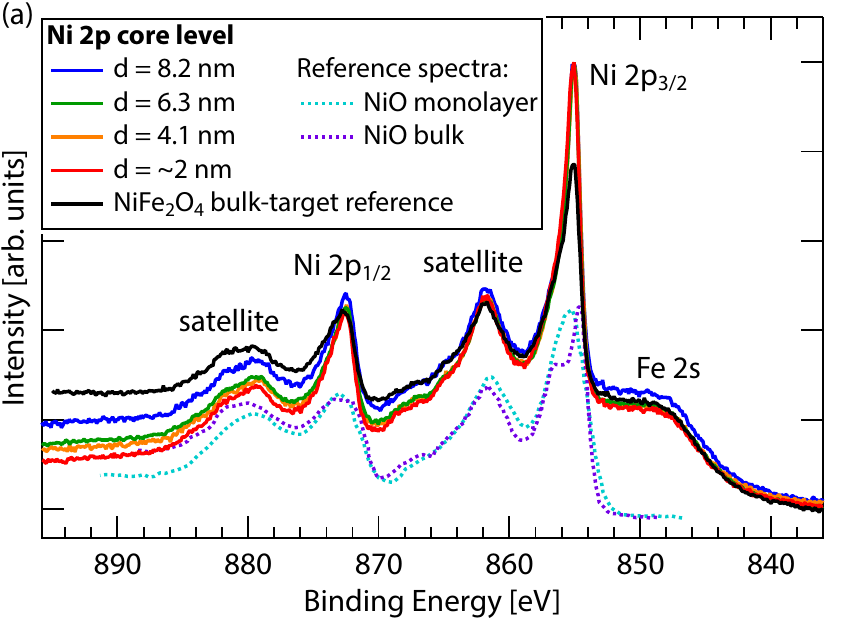}
	\includegraphics{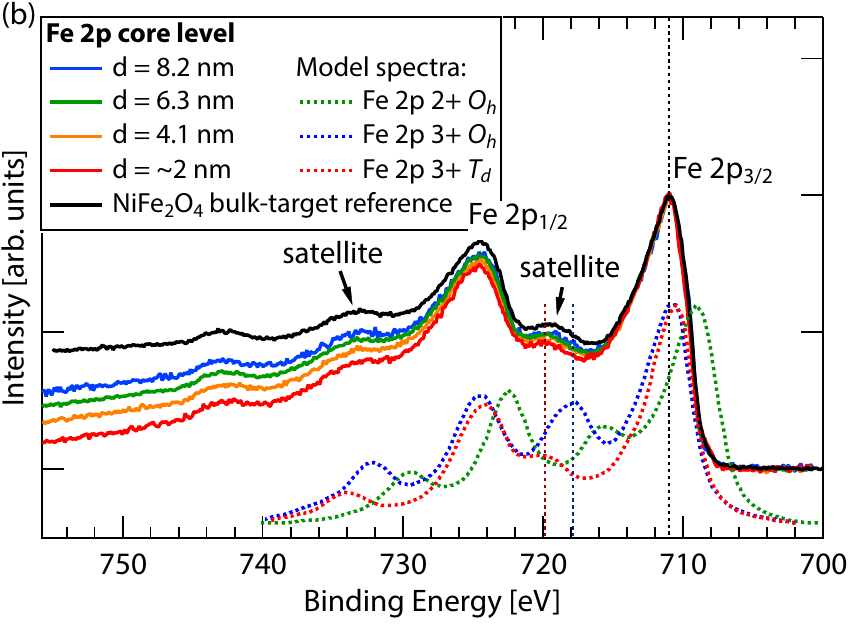}
	\vspace*{-6mm}
	\caption{HAXPES spectra of NiFe\textsubscript{2}O\textsubscript{4} films with varying film thickness recorded at a photon energy of $h\nu = \SI{4}{\kilo\electronvolt}$.
	(a) Ni 2p core level spectra and references for one monolayer NiO and NiO bulk \cite{alders_nonlocal_1996}. (b) Fe 2p core level spectra in comparison to model spectra taken from  \cite{fujii_situ_1999}.}
	\label{fig:haxpes-ni+fe2p}
	\end{figure}

	In a first step, we need to clarify whether the chemical properties of NiFe\textsubscript{2}O\textsubscript{4} differ for bulk-like and ultrathin films. HAXPES measurements have been performed to quantify the valence states of each cation species. In contrast to soft X-ray photoemission, HAXPES allows us to identify these properties not only at the surface but with bulk sensitivity. The increased information depth even allows us to record reference spectra of the pressed NiFe\textsubscript{2}O\textsubscript{4} powder used as bulk-target for PLD deposition, which do not posses a flat surface as typically required for low-energy photoemission experiments.

	Figure \ref{fig:haxpes-ni+fe2p} plots the Ni 2p and Fe 2p core level spectra for NiFe\textsubscript{2}O\textsubscript{4} films of \SI{8}{\nano\metre} to \SI{2}{\nano\metre}, and compares them to the bulk reference. In Figure \ref{fig:haxpes-ni+fe2p}(a), all spectra of the Ni 2p core level display the Ni 2p\textsubscript{3/2} and Ni 2p\textsubscript{1/2} peaks at a binding energy of \SI{855.1}{\electronvolt} and \SI{872.4}{\electronvolt} respectively, without a chemical shift relative to the bulk material. The two main peaks are both accompanied by satellite peaks at 7 eV above their binding energies and overlap with the Fe 2s core level at lower energies. The shape of all spectra is comparable to that of a single monolayer of NiO \cite{alders_nonlocal_1996}, in particular there is no shoulder visible at the high energy side of Ni 2p\textsubscript{3/2}. The occurrence of such a shoulder $\sim$\SI{1.5}{\electronvolt} above the 2p\textsubscript{3/2} peak (see NiO bulk reference in Fig. \ref{fig:haxpes-ni+fe2p}(a) for comparison) has been theoretically described by a screening effect, that emerges from electrons not originating from the oxygen orbitals around the excited Ni cation, but from adjacent NiO\textsubscript{6} clusters \cite{van_veenendaal_nonlocal_1993}. The HAXPES experiment thus confirms, that no NiO clusters have formed within the NiFe\textsubscript{2}O\textsubscript{4} films. Moreover, the spectra do not show any contribution of metallic Ni\textsuperscript{0}, which would peak at around 852.8 eV. We therefore conclude, that the NiFe\textsubscript{2}O\textsubscript{4} films contain completely oxidized and homogeneously distributed Ni\textsuperscript{2+} cations only without any NiO cluster formation.

	Figure \ref{fig:haxpes-ni+fe2p}(b) depicts the HAXPES data of the Fe 2p core levels from all NiFe\textsubscript{2}O\textsubscript{4} samples with $d = 8 - \SI{2}{\nano\metre}$. For comparison, also model spectra of Fe cations in the inverse spinel structure of magnetite (Fe\textsubscript{3}O\textsubscript{4}) are given (reproduced from Ref. \cite{fujii_situ_1999}). These spectra have been calculated individually for the different possible Fe cation lattice site occupancies ($O_h$, $T_d$) and valencies (2+, 3+). The Fe 2p\textsubscript{3/2} and 2p\textsubscript{1/2} core levels are peaking at binding energies of \SI{710.9}{\electronvolt} and \SI{724.4}{\electronvolt}, in agreement with the spectrum of the NiFe\textsubscript{2}O\textsubscript{4} bulk reference and consistent with literature \cite{biesinger_resolving_2011}. The formation of under-oxidized Fe\textsuperscript{2+} ions during film growth would result in a characteristic shoulder at the low-energy side of the Fe 2p\textsubscript{3/2} peak, due to a chemical energy shift (as observable in the Fe\textsuperscript{2+} $O_h$ reference). All measured spectra coincide with the bulk reference sample, thus confirming that the NiFe\textsubscript{2}O\textsubscript{4} films consist of fully oxidized Fe\textsuperscript{3+} cations and that the amount of underoxidized Fe\textsuperscript{2+} cations is below the detection limit.

	Comparing the Fe\textsuperscript{2+} $O_h$, Fe\textsuperscript{3+} $O_h$ and Fe\textsuperscript{3+} $T_d$ model spectra reveals, that the main peak binding energies are sensitive to the oxidation state, but not to the atomic site occupancy. In contrast, the Fe 2p\textsubscript{3/2} satellite observable between the spin-orbit split Fe 2p peaks is caused by a screening effect of the surrounding oxygen ions and deviates significantly for $T_d$ and $O_h$ cation coordination. Thus, its shape and binding energy position can serve as a fingerprint for the chemical state of different iron oxides and the cationic lattice site occupancies \cite{fujii_situ_1999}. A complete inversion to the normal spinel structure would shift the satellite's spectral weight to lower binding energies by about 0.8 eV. Since both shape and energy position of the thin film samples satellite peaks perfectly match that of the NiFe\textsubscript{2}O\textsubscript{4} bulk spectrum, we conclude, that the Fe\textsuperscript{3+} cations occupy the bulk lattice sites -- without any sign for a cationic inversion from the inverse to the normal spinel structure in the binding energy resolution limit of the performed HAXPES experiment.

	In summary, both the Ni 2p and Fe 2p spectra are comparable to the spectrum of bulk material for all film thicknesses, and reveal that the chemical composition of the bulk material is well reproduced in the ultrathin NiFe\textsubscript{2}O\textsubscript{4} films. The Fe 2p\textsubscript{3/2} satellite gives no hint for a cationic inversion in ultrathin NiFe\textsubscript{2}O\textsubscript{4}. In order to rule out also any smaller effect, we investigate the spatial cationic distribution by further spectroscopic means.

	\subsection{XANES}
	%%%%%%%%%%%%%%%%%%%%%%%%%%%%%%%%%%%%%%%%%%%
	%%%%%%%%%%%%%%%%%%%%%%%%%%%%%%%%%%%%%%%%%%%

	\begin{figure}
	\includegraphics{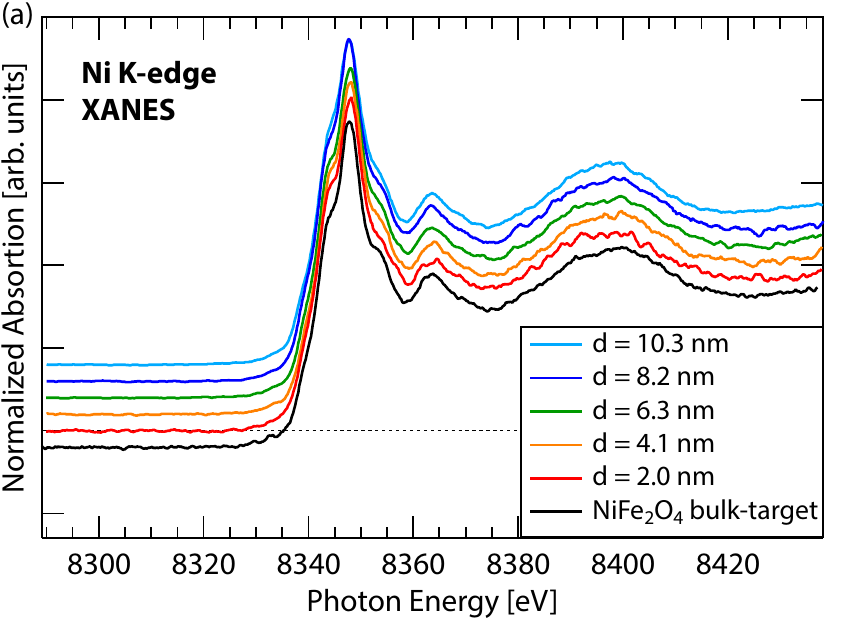}
	\includegraphics{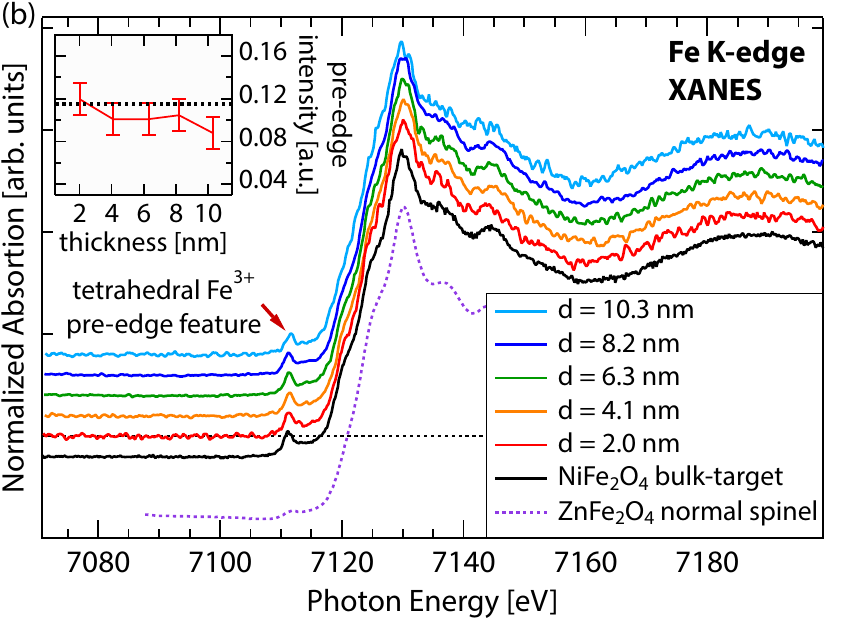}
	\vspace*{-6mm}
	\caption{XANES spectra of (a) the the Ni \textit{K}-edge and (b) the Fe \textit{K}-edge from NiFe\textsubscript{2}O\textsubscript{4} films with varying film thickness. For comparison a Fe K-edge spectrum of Zn ferrite, which exhibits the normal spinel structure, is plotted (reproduced from \cite{matsumoto_site-specific_2000}). The inset in Fig. \ref{fig:xanes-fe}(b) shows the integrated spectral weight of the Fe \textit{K}-edge pre-edge feature in dependence of the film thickness, where the dotted line represents the bulk value.}
	\label{fig:xanes-fe}
	\end{figure}

	To gain precise information on the spatial cationic distribution in the NiFe\textsubscript{2}O\textsubscript{4} thin films, we recorded XANES spectra of the Fe and Ni \textit{K}-edge. Since the fine structure above the absorption edge is dominated by multiple scattering with the surrounding atoms of the investigated cation species, XANES is very sensitive to the distribution of the oxygen anions around the cation. Thus, a cationic inversion - for which the local site occupancy changes from tetrahedral to octahedral, or vice versa - considerably modifies the shape of the spectral fine structure. 

	The absorption spectra of the NiFe\textsubscript{2}O\textsubscript{4} film samples are recorded by fluorescence yield, thus the measured data probes the bulk-like film properties. Figure \ref{fig:xanes-fe}(b) shows the XANES spectra of the Fe \textit{K}-edge for NiFe\textsubscript{2}O\textsubscript{4} films down to \SI{2}{\nano\metre} and a bulk material reference spectrum. All spectra show a pre-edge feature at 7111 eV, which in case of the spinel structure is observable for cations in a $T_d$ symmetry only. While the main absorption line is caused by a dipole transition from the 1s to the empty 4p orbital, the pre-edge structures in transition metal oxides are assigned to quadrupole transitions to the empty 3d states, and thus are only very weak \cite{groot_1s_2009}. If the inversion symmetry of the transition metal cation is broken, the local 3d and 4p wavefunctions of the cation hybridize, and in turn dipole transitions into this orbital become allowed, leading to an increased weight of the pre-edge feature. In the spinel structure a broken symmetry is given for cations on $T_d$, but not on $O_h$ sites.

	XANES studies of the Fe \textit{K}-edge of various spinels clearly show a sharp pre-edge for all materials exhibiting the inverse spinel structure, where Fe cations are situated on $T_d$ sites. In contrast, the spectra of compounds featuring the normal spinel structure, in which Fe cations solely occupy $O_h$ sites, only show a weak broad feature  \cite{matsumoto_site-specific_2000}. In Fig. \ref{fig:xanes-fe}(b), this is exemplary shown by a XANES reference spectrum of the normal spinel ZnFe\textsubscript{2}O\textsubscript{4} (reproduced from  \cite{matsumoto_site-specific_2000}).

	The normalized pre-edge intensity can be quantitatively correlated to the local site symmetry of the investigated cation species  \cite{wilke_oxidation_2001}. By monitoring the Fe \textit{K}-edge pre-edge intensity of the NiFe\textsubscript{2}O\textsubscript{4} samples, no intensity changes are resolvable between the various film thicknesses and also not in comparison to the bulk reference sample. We thus can conclude once more, that the ultrathin films do not undergo a cationic inversion, but remain in the bulk-like cationic distribution of the inverse spinel lattice. This is supported by the Ni \textit{K}-edge spectra (Fig. \ref{fig:xanes-fe}(a)), which also show no sign of an emerging pre-edge feature, characteristic for Ni cations on $T_d$ sites.

	Focussing on the main Fe \textit{K}-edge in Fig. \ref{fig:xanes-fe}(b), a chemical shift is expected for valency changes. A shift of about 5 eV between Fe\textsuperscript{2+} and Fe\textsuperscript{3+} for octahedrally coordinated iron oxides was observed previously \cite{sasaki_fe2+_1995}. A comparison of XANES spectra from bulk Fe\textsubscript{3}O\textsubscript{4} with NiFe\textsubscript{2}O\textsubscript{4}, for which Fe\textsuperscript{2+} cations are replaced by Ni\textsuperscript{2+}, reveals a chemical shift of about 3 eV, that has been explained by the missing Fe\textsuperscript{2+} ions \cite{saito_site-_1999}. This energy shift has also been observed in other ferrites, where the Fe\textsuperscript{2+} cations are substituted by a different cation species \cite{matsumoto_site-specific_2000}. In all cases, the Fe-\textit{K}-edge of the Fe\textsuperscript{2+}-compounds was situated at lower binding energies. In our case, we observe no chemical shift of the main-edge across all film thicknesses, thus again supporting that no modification in the oxidation state of the Fe cations occurs, fully consistent with our HAXPES results.

	The results of this in-depth XANES pre-edge analysis clearly reveal that NiFe\textsubscript{2}O\textsubscript{4} films grow in the inverse spinel structure independent of their film thickness. Moreover, the position of the main \textit{K}-edges confirms, that the Fe and Ni cations in all samples are present in a bulk-like valency for all film thicknesses.

	\subsection{XMCD}
	%%%%%%%%%%%%%%%%%%%%%%%%%%%%%%%%%%%%%%%%%%%
	%%%%%%%%%%%%%%%%%%%%%%%%%%%%%%%%%%%%%%%%%%%

	\begin{figure}
	\includegraphics{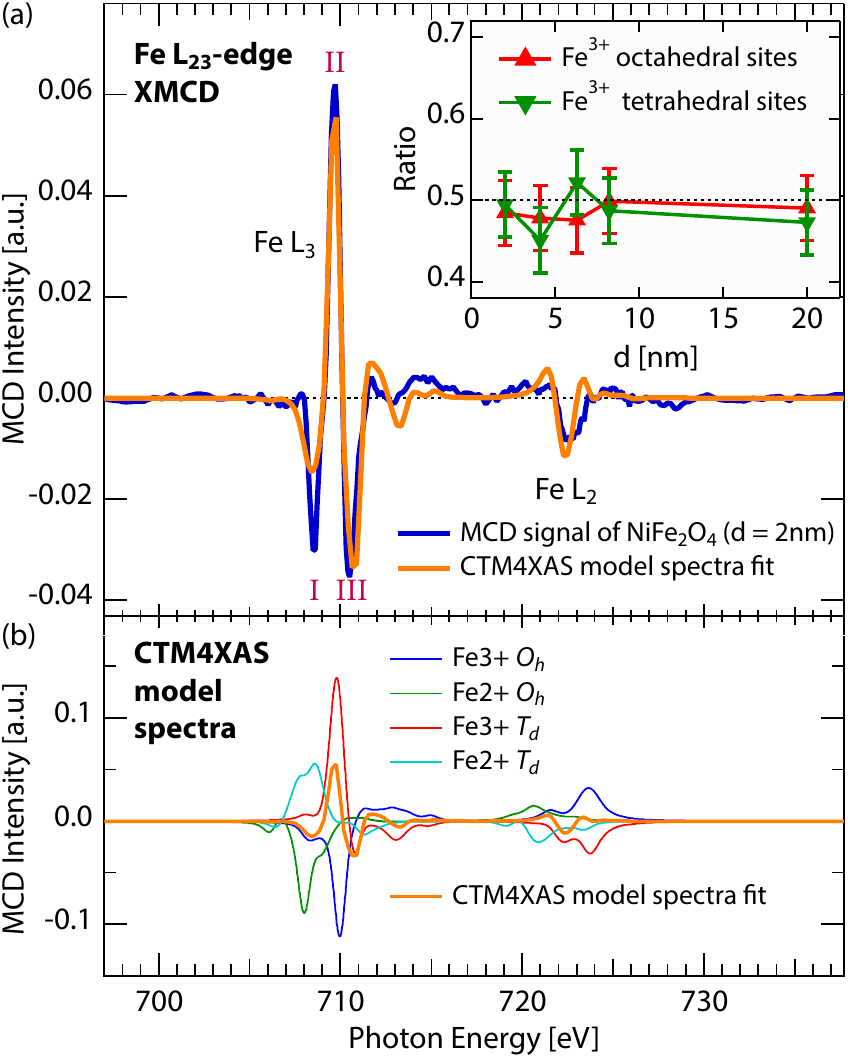}
	\vspace*{-6mm}
	\caption{Experimental XMCD spectrum from the Fe \textit{L}\textsubscript{2,3}-edge of the \SI{2}{\nano\metre} thick NiFe\textsubscript{2}O\textsubscript{4} film and the corresponding fit. The resulting lattice site occupancy for various film thicknesses is depicted in the inset.}
	\label{fig:xmcd-fe}
	\end{figure}

	In a last step, we analyse the XMCD asymmetry signal to quantitatively determine the cationic distribution across the spinel lattice sites. The XMCD asymmetry spectra are element-specific and sensitively influenced by the valency, the local lattice site symmetry and the magnetic ordering of the investigated cation species. The spectral details reflect the superposition of cations occupying $T_d$ or $O_h$ sites with either divalent or trivalent valency, respectively. Hereby, each configuration has its own characteristic MCD spectrum, which serves as a fingerprint for the certain atomic and geometric configuration. We thus modelled those four XMCD spectra, which allows us to fit them as a linear combination to the experimental data and to quantify the fraction of each configuration.

	We investigated XMCD asymmetry spectra of the Fe \textit{L}\textsubscript{2,3}-edge to identify any changes in the distribution of Fe cations between $T_d$ and $O_h$ lattice sites for NiFe\textsubscript{2}O\textsubscript{4} films with varying thickness. Site- and valency-specific Fe \textit{L}\textsubscript{2,3}-edge XMCD spectra were computed by LFM calculations utilizing the software CTM4XAS \cite{stavitski_ctm4xas_2010}, which are presented in Fig. \ref{fig:xmcd-fe}(b). Due to the antiferromagnetic alignment of the cation spins between $T_d$ and $O_h$ sites, their asymmetry signals are of opposite sign. Consequently, these signals mainly cancel out in the observable sum asymmetry signal, leaving the resulting difference spectrum extremely sensitive to subtle changes in the cationic distribution.

	Fig. \ref{fig:xmcd-fe}(a) exemplary shows the MCD spectrum of the Fe \textit{L}\textsubscript{2,3}-edge for the \SI{2}{\nano\metre} thick NiFe\textsubscript{2}O\textsubscript{4} film with the corresponding fit. The \textit{L}\textsubscript{3}-edge exhibits a pronounced -/+/- asymmetry structure, caused by the antiparallel oriented Fe moments. The positive (+) peak at \SI{709.7}{\electronvolt} (II) is dominated by tetrahedral Fe\textsuperscript{3+} and the high energy negative (-) peak at \SI{710.5}{\electronvolt} (III) by octahedral Fe\textsuperscript{3+} cations. We note, that the first negative peak at \SI{708.5}{\electronvolt} (I) is enhanced in comparison to the model and to the reference data \cite{pattrick_cation_2002}, thus indicating that a fraction of Fe\textsuperscript{2+} cations of about $\approx \SI{2}{\percent}$ is existent at the film surface. However, the result gives no indication for a cationic inversion of the film, which would result in a decrease of the positive peak (II) and strong enhancement of the high energy negative peak (III). These results are also observed for all other investigated NiFe\textsubscript{2}O\textsubscript{4} film thicknesses, which give no clue for an increased octahedral Fe\textsuperscript{3+} fraction, as would be characteristic for a cationic inversion to the normal spinel structure. This finding is in perfect agreement with electronic structure calculations, which find the fully inverse spinel lattice to be the ground state of bulk NiFe\textsubscript{2}O\textsubscript{4} \cite{szotek_electronic_2006}.

	Since the XMCD spectra are recorded in TEY mode, the experiment probes the uppermost 2-3 \SI{}{\nano\metre} of material. Complemented by the bulk-sensitive HAXPES and XANES techniques, the analysis yields a consistent picture of the stoichiometry, valency and cationic distribution of the NiFe\textsubscript{2}O\textsubscript{4} thin films. In particular, we find that the cationic site occupancy always belongs to that of an inverse spinel lattice -- this result is found both at the NiFe\textsubscript{2}O\textsubscript{4} surface and in the bulk volume. This striking consistency provides clear evidence for the absence of a cationic inversion in NiFe\textsubscript{2}O\textsubscript{4} in the crossover to the ultrathin film limit, and thus rules out this mechanism as the origin of the observed enhanced $M_S$ in ultrathin NiFe\textsubscript{2}O\textsubscript{4} films.

%%%%%%%%%%%%%%%%%%%%%%%%%%%%%%%%%%%%%%%%%%%
%	Summary
%%%%%%%%%%%%%%%%%%%%%%%%%%%%%%%%%%%%%%%%%%%
\section{Summary}\label{sec:summary}

	In summary, we have investigated single-crystalline NiFe\textsubscript{2}O\textsubscript{4} thin films grown cube-on-cube on Nb-doped SrTiO\textsubscript{3} (001) substrates, with thicknesses scaling down from 20 -- \SI{2}{\nano\metre}. In this crossover to the ultrathin film limit, we focussed on the impact of reduced dimensionality on the structural, electronic and magnetic NiFe\textsubscript{2}O\textsubscript{4} properties. Foremost, we observed an enhanced saturation magnetization $M_S$ in ultrathin NiFe\textsubscript{2}O\textsubscript{4} films. Despite the substrate-induced compressive in-plane strain, a reduced out-of-plane NiFe\textsubscript{2}O\textsubscript{4} lattice constant is found, implying that a reduction of the unit-cell volume is energetically favourable. In order to investigate the cationic distribution in the NiFe\textsubscript{2}O\textsubscript{4} thin films,  complementing bulk- and surface-sensitive analyses using HAXPES, XANES and XMCD spectroscopy techniques have been performed, and special attention was paid to the element-specific cation valencies and -coordinations. We find a bulk-like inverse spinel structure being present in all samples -- independent of the NiFe\textsubscript{2}O\textsubscript{4} film thickness. Thereby, our results consistently reveal the absence of a cationic inversion from the inverse to the normal spinel structure, as was so far held responsible for an enhanced $M_S$ in ultrathin spinels. From our experimental results we thus propose an auxetic behavior, i.e. a structural unit cell reduction, being a possible mechanism for a stronger interatomic electronic localization in ultrathin NiFe\textsubscript{2}O\textsubscript{4} films. Theoretical calculations that shall further elucidate possible underlying physical mechanisms are currently underway.

%%%%%%%%%%%%%%%%%%%%%%%%%%%%%%%%%%%%%%%%%%%
%	Acknowledgement and bibliography      %
%%%%%%%%%%%%%%%%%%%%%%%%%%%%%%%%%%%%%%%%%%%
\begin{acknowledgments}
	We acknowledge experimental support during beamtimes by B. Zijlstra and C. Caspers. We thank R. Dittmann for providing the PLD setup at FZJ. This work has been funded by the Helmholtz Association under Grant HGF-NG-811.

\end{acknowledgments}

\end{document}